\documentstyle[amstex,12pt]{article} 

\newcommand{\be}{\begin{equation}}
\newcommand{\ee}{\end{equation}}
\newcommand{\bea}{\begin{eqnarray}}
\newcommand{\eea}{\end{eqnarray}}
\newcommand{\cp}{{\Bbb C \Bbb P}}
\newcommand{\oo}{{\cal O}}
\newcommand{\M}{{\bf M}}
\newcommand{\I}{{\bf I}}
\newcommand{\J}{{\bf J}}
\newcommand{\K}{{\bf K}}
\newcommand{\T}{{\bf T}}
\newcommand{\G}{{\Gamma}}

\def\Bbb{\bf} 

\input epsf

\begin{document}
\begin{titlepage}
\begin{flushright}
ITP-SB-95-54\\
hep-th/9512075\\
December 1995
\end{flushright}
\bigskip
\Large
\begin{center}
\bf{Supersymmetric $\,\sigma$-models, twistors, and the Atiyah-Hitchin
metric}\\

\bigskip

\normalsize
Ivan T. Ivanov\footnote{email: iti@@insti.physics.sunysb.edu}
and M.\ Ro\v cek\footnote{email: rocek@@insti.physics.sunysb.edu}\\
{\it Institute for Theoretical Physics\\
State University of New York\\
Stony Brook, NY 11794-3840 USA\\}

\end{center}
\vspace{1.0cm}
\begin{abstract}
The Legendre transform and its generalizations, originally found in
supersymmetric $\sigma$-models, are techniques that can be used to give
{\em local} constructions of hyperk\"ahler metrics. We give a twistor
space interpretation to the generalizations of the Legendre transform
construction. The Atiyah-Hitchin metric
on the moduli space of two monopoles is used as a detailed example.
\end{abstract}
\end{titlepage}
\eject
\section{Introduction.}
Some time ago, we discoverd several constructions of hyperk\"ahler
metrics by studying extended supersymmetric nonlinear sigma-models
\cite{lr1,hklr,klr,lr2}. In the hyperk\" ahler quotient framework, the
topology of the manifold and the completeness of the metric are
accessible. On the other hand, the Legendre transform construction and
its generalizations are purely local. They can be considered as changes
of variables that linearize the Monge-Amp\`ere equations governing
hyperk\" ahler metrics. In some cases, the Legendre transform
construction {\em has} been used to find complete metrics. A twistor
space interpretation has also been worked out. In this paper, we
consider the simplest generalization of the Legendre transform
construction \cite{lr2}. We reformulate it using twistors and empasize
the special properties of the twistor spaces used. Throughout this
paper we consider only the four dimensional case to avoid congesting
the notation, but all the techniques are easily generalized to higher
dimensions. As the main example we show that the hyperk\"ahler metric
in the moduli space of two monopoles \cite{AH} can be obtained by a
generalized Legendre transform construction from a single simple
function. Using this example we also propose various ingredients in the
construction of new complete metrics.

In section two we introduce the generalized Legendre transform
construction from \cite{lr2}. Section three contains a brief and, we
hope, readable outline of the basics of twistor space. In section
four, the twistor interpretation of the generalized Legendre transform
construction is worked out. A careful treatment of the singularities of
the functions involved is essential. Section five briefly reconsiders
the twistor construction of $A_k \; ALE$-metrics from a new point of
view; in particular, the singularities that arise in this construction,
which were regarded in \cite{hklr} as something of an embarrassment, are
understood and shown to play a crucial role. Section six is devoted to the
Atiyah-Hitchin metric in the framework of the generalized Legendre
transform construction. In the final section we notice that many of the
complete metrics we have discussed in the paper are built from a few ``flat
space'' ingredients. We propose more complete metrics based on these.

\section{Generalized Legendre transform construction.}
To set the stage we review the generalized
Legendre transform construction of hyperk\"ahler metrics \cite{lr2}.
Consider $\eta (\zeta)= \sum_{a=0}^{2n} w_a \zeta^a$ to be a
polynomial of order $2n$ in $\zeta$, where $\zeta$ is the coordinate on
the Riemann sphere $\cp_1 = S^2$. This polynomial should obey a reality
condition: $\overline{\eta (\zeta)} = (-1)^{n} \bar{\zeta}^{2 n} \eta
(-1/\bar{\zeta})$, which implies $w_{2n-a}=(-1)^{n+a}\bar w_a$. The
K\"ahler potential is constructed from one function of two variables
$G(\eta (\zeta),\zeta)$. Consider the function $F$
\bea
F (w_a) = \frac1{2\pi i} \oint \limits_C \frac{d\zeta}{\zeta^2}
G(\eta,\zeta)
\eea
where $C$ is some appropriately chosen contour.
$F$ obeys a system of linear differential equations
\bea
F_{ab} = F_{a+c,b-c}, \ \text{where} \ F_{ab}\equiv\frac{\partial^2
F}{\partial w_a \partial w_b}\ .
\eea
Let $z=w_0, v=w_1$; then $w_{2n}=(-1)^n\bar z$ and $w_{2n-1}= -(-1)^n
\bar v$. The K\"ahler potential is constructed from the function
$F(w_a)$ by performing a two dimensional Legendre transform with respect
to $v$ and $\bar v$, and extremizing with respect to the
remaining coordinates $\{w_{a'}, 2 \leq a' \leq 2n-2\}$.
\bea
K(z,\bar z,u,\bar u) = F(z,\bar z,v,\bar v,w_{a'}) - uv - \bar u\bar
v\ ,
\eea
where
\bea
 v = v(z,\bar z,u,\bar u)\ ,\ w_{a'} = w_{a'}(z,\bar z,u,\bar u)\ ,
\eea
are solutions to the equations
\bea
F_v = u\ ,\ F_{a'} = 0\ , \ 2 \leq a' \leq 2n-2\ .
\eea
The metric is obtained as usual from the K\"ahler potential and can be
expressed in terms of second derivatives of the function $F(w_a)$. We
refer to \cite{lr2} for the explicit equations.

\section{Brief review of twistor approach.}
We start with very brief review of twistor theory for hyperk\"ahler
manifolds (for details and proofs see \cite{hklr}).
A hyperk\"ahler manifold $\M$ is a Riemannian manifold which is
K\"ahler with
respect to three complex structures \I, \J, and \K\ that
obey the quaternionic algebra relations
\bea
\I^2 = \J^2 = \K^2 = -1\ , \ \I\J = -
\J\I = \K\ , \ \text{etc.}
\eea
The K\"ahlerian conditions imply that the metric $g$ on $\M$ is
hermitian with respect to the complex structures and the complex
structures are covariantly constant with respect to the Levi-Civita
connection: for arbitrary vector fields {\bf X}, {\bf Y}
\bea
g ({\bf IX},{\bf IY}) = g ({\bf X},{\bf Y}) \ ,\ \nabla \I = 0\ ,\
\text{etc.}
\eea
Clearly, if $(a,b,c)$ is a unit vector in ${\Bbb R}^3$, the properties
of \I, \J, and \K\ imply that
$\I (a,b,c) = a \I + b \J + c \K$ is another
complex structure compatible with the metric and the connection. Thus,
on a hyperk\"ahler manifold, there is a two-sphere's worth of K\"ahler
structures. The twistor space $\T$ of the hyperk\"ahler manifold
$\M$ is then constructed so as to incorporate all the
complex structures together on a larger complex manifold \cite{hklr}.

As a manifold the twistor space $\T$ is just the product $\M\times
S^2$. Accordingly, the tangent space to $\T$ splits as the direct
sum $\T_m \oplus \T_{\zeta}$. To see that $\T$ is a complex
manifold, we first consider a complex structure on $S^2$. The
Riemann sphere $S^2$ considered as the complex projective line $\cp_1$
is a complex manifold obtained by patching together two copies $U, \;
\tilde U$ of the complex plane $\Bbb C$ with coordinates $\zeta, \;
\tilde\zeta$ related by $\tilde \zeta = \zeta^{-1}$ on $U \cap \tilde
U$. We will use this model of $\cp_1$ throughout the paper and refer
to $\zeta$ as the standard coordinate on it. Define a almost complex
structure $\I(\zeta)$ on $\T$ by
\bea
{\bf \b I} = \left( \I (\zeta), \; I_0 \right) =\left(
\frac{1-\zeta \bar\zeta}{1+\zeta\bar\zeta} \, \I +
\frac{\zeta + \bar\zeta}{1+\zeta\bar\zeta} \, \J + i
\frac{\zeta - \bar\zeta}{1+\zeta\bar\zeta} \, \K \; , \; I_0
\right)\ .
\label{mama}
\eea
Here the two operators $\I (\zeta)$ and $I_0$ act on $\T_m$ and
$\T_{\zeta}$ respectively. $I_0$ is the standard complex structure on
$\cp_1$ as described above.

It is easy to prove that ${\bf \b I}$ is an {\em integrable} almost
complex structure and so $\T$ is a complex manifold and admits
complex coordinates. Moreover the projection $\T \to \cp_1$ is
holomorphic and each copy $(m,\cp_1)$ (where $m$ is a point on $\M$)
of the projective line is a holomorphic section of this projection. These
sections are called twistor lines.

Another key ingredient in the twistor space approach is a certain
holomorphic symplectic form. For a K\"ahler manifold the metric $g$
and the complex structure \J\ can be combined into a symplectic form
$\omega ({\bf X},{\bf Y})\equiv g ({\bf JX},{\bf Y})$ called the
K\"ahler form. For a hyperk\"ahler manifold, there are three K\"ahler
forms $\omega_1,\ \omega_2,\ \text{and}\ \omega_3$ corresponding to
\I, \J, and \K\ respectively. Let $z_1 \; \text{and} \; z_2$ be
(local) holomorphic coordinates on $\M$ with respect to the complex
structure \I, {\it i.e.,\/} $\I d z_1 = i d z_1,\
\I d z_2 = i d z_2$. It is easy to prove that one can choose the
coordinates $z_1,z_2$ such that $d z_1\wedge d z_2 = (\omega_2 + i
\omega_3)/2$. Thus the two-form $\omega_+=(\omega_2 + i \omega_3)/2$
is a holomorphic (2,0)-form with respect to the complex structure \I. One
can extend $\omega_+$ to a holomorphic form on \T\ by defining:
\bea
\omega_h(\zeta)({\bf X},{\bf Y})\equiv\omega_+({\bf X+\zeta K X},\; {\bf
Y+\zeta K Y})\ .
\eea
The Clifford algebra of the complex stuctures guarantees that $\omega_h
(\zeta)$ is holomorphic (2,0)-form for $\I (\zeta)$ and can be written out as
\bea
2 \omega_h (\zeta) = (\omega_2 + i \omega_3) + 2 \zeta \omega_1 -
\zeta^2 (\omega_2 - i \omega_3)\ .
\label{holform}
\eea
Note that $\omega_h (\zeta)$ is quadratic in $\zeta$;
this is expressed mathematically by saying that $\omega_h
(\zeta)$ is section of $\bigwedge^{2} \M$ ({\it i.e.,} a 2-form)
twisted by $\oo (2)$. Here $\oo (2)$ is the line bundle over $\cp_1$ of
degree two. In general, by $\oo (n)$ we denote the line bundle over
$\cp_1$ of degree $n$. If $\zeta$ is the standard coordinate on
$\cp_1$, the sections of the bundle $\oo (n)$ can be identified with
polynomials of degree $n$: $a_n \zeta^n + \dots + a_1 \zeta +a_0$.

The last ingredient in the twistor space picture is a map $ \tau :
\M \times S^2 \to \M \times S^2$ called a {\it real structure}. It is
defined by
\bea
\tau (m, \zeta) = \left( m, - \frac{1}{\bar \zeta} \right)\ .
\eea
Thus $\tau$ is the antipodal map on the $S^2$ factor, and from
(\ref{mama}) it takes the complex structure ${\bf \b I}$ to its
conjugate $- {\bf \b I}$. It codes ordinary complex conjugation. 
All the holomorphic data given is compatible with the real
structure composed with complex conjugation (see, {\it e.g.}, eq.\ 
\ref{real}).

This construction is invertible, and given a twistor space with all the
properties above, we can reconstruct \M\ and the hyperk\"ahler metric
on it. We will not go into any further details, and merely refer the
interested reader to \cite{hklr} and \cite{BESSE}.

In practice it is
quite difficult to obtain hyperk\"ahler metrics from twistor spaces
even if the data needed on them is essentially free, because some steps
in the inverse construction (in particular finding the twistor lines)
are quite involved. The relatively easy case of hyperk\"ahler metrics
with tri-holomorphic isometries is treated thoroughly in \cite{hklr}
(see also \cite {ALA}) and sometimes referred to as the Legendre
transform construction. In the next section we present a generalization
that works for metrics with special twistor spaces.

\section{Twistor point of view on the generalized Legendre transform
construction.}

By construction, the twistor space \T\ of the hyperk\"ahler manifold
\M\ has a projection onto $\cp_1$, and the fibers of this bundle are
copies of \M. We consider twistor spaces which also have
(intermediate) holomorphic projections onto line bundles over $\cp_1$
of positive even degree.
\bea
\T \to \oo (2n) \to \cp_1\ .
\eea

This requirement is equivalent to the existence of a $\I
(\zeta)$-holomorphic coordinate $\eta (\zeta)$ on \M, such that $\eta
(\zeta)$ is polynomial of order $2n$. An additional condition we
impose is that $\eta (\zeta)$ is real under the real structure $\tau$
\bea
\overline{\eta (\zeta)} =
(-1)^n \bar{\zeta}^{2n} \eta (-1/\bar \zeta)\ .
\label{real}
\eea
The case $n=1$ applies to metrics with triholomorphic isometry.

Our model of $\cp_1$ was build by two patches $U$ and $\tilde U$ with
local coordinates $\zeta$ and $\tilde \zeta$ related by $\tilde \zeta =
\zeta^{-1}$ on the intersection $U \cap \tilde U$. The polynomial
expression for the complex coordinate $\eta (\zeta) = \sum_{a=0}^{2n}
w_a \zeta^a$ is good only on the patch $U$. On $\tilde U$ the
coordinate is $\tilde \eta (\zeta^{-1})$ and on $U \cap \tilde U$ the
two are related by
\bea
\tilde \eta = \zeta^{-2n} \; \eta\ .
\label{eta}
\eea
Let $\chi (\zeta)$ ($\tilde \chi (\zeta)$) be the second
holomorphic coordinate on \M\ over $U$ ($\tilde U$), respectively. 
The holomorphic form $\omega_h(\zeta)$ can be evaluated and 
compared on the two patches
\bea
\omega_h = d \eta \wedge d \chi =
\zeta^2 d \tilde \eta \wedge d \tilde \chi \ .
\label{omega}
\eea
The second equality follows because $\omega_h$ is a two form twisted by
$\oo (2)$. From (\ref{eta}) and (\ref{omega}) we see that
the expressions for the second holomorphic coordinate on the two
patches $U$ and $\tilde U$ are connected by
\bea
\tilde \chi = \zeta^{2n-2} \; \chi - f (\eta,\zeta)\ .
\label{important}
\eea
Here $f (\eta,\zeta)$ is arbitrary function of $\eta$ and $\zeta$. Even if
$f (\eta,\zeta) = 0$, we can not conclude that $\chi (\zeta)$ is
a section of $\oo (2-2n)$, since it can have poles or cuts or both as a
function of $\zeta$. As we shall see, the singularities of $\chi
(\zeta)$ play a crucial important role in determining the metric on $\M$.

We now expand the coordinates $\eta (\zeta)$ and $\chi (\zeta)$ in
Taylor series around the point $\zeta = 0$.\footnote{We assume that the
singularities of $\chi$ are away from $\zeta=0$.}
\begin{align}
\begin{split}
\eta (\zeta) = z + v \zeta &+ w_2 \zeta^2 + \dots + \\
 + & w_{2n-2} \zeta^{2n-2} +
(-1)^{n-1} \bar v \zeta^{2n-1} + (-1)^n \bar z \zeta^{2n}
\end{split} \\
\chi (\zeta) = u + t \zeta &+ O (\zeta^2)\ .
\end{align}
Note that the particular expansion of $\eta (\zeta)$ is guided by the
reality condition (\ref{real}), and $z$ and $u$ are complex
coordinates for the complex structure at $\zeta = 0$, namely \I. The
expansion of $\tilde \eta (\zeta)$ around $\zeta = \infty$ follows from
(\ref{eta}) and for the expansion of $\tilde \chi$ we write
\bea
\tilde \chi (\zeta^{-1}) = \tilde u + \tilde t \zeta^{-1}
+ O (\zeta^{-2})\ .
\eea
Next, following \cite{hklr}, we compare the coefficients for different
powers of $\zeta$ in (\ref{important}). Consider a integral along a
small contour around the point $\zeta = \infty$ of both sides of
equation (\ref{important})
\bea
\frac{1}{2 \pi i} \oint \limits_\infty\frac{d \zeta}{\zeta^{m}} \;
\tilde \chi = \frac{1}{2 \pi i} \oint \limits_\infty\frac{d \zeta}
{\zeta^{m+2-2n}} \; \chi - \frac{1}{2 \pi i} \oint \limits_\infty
\frac{d \zeta}{\zeta^m} \; f (\eta, \zeta)\ .
\label{round}
\eea
The integral on the left hand side will pick up the
appropriate coefficient in the Taylor expansion. To select the
appropriate coefficient of the $\chi$ expansion, however, the contour
should be deformed to a contour around $\zeta = 0$. In the process
we will pick contributions from the singularities of $\chi$ at points
$\zeta \neq \infty$. We assume that the corresponding terms can be
expressed in terms of a contour integral of a function
$f'(\eta,\zeta)$. This assumption is satisfied for a host of
examples and amounts to saying that the twistor lines are specified 
by a choice of the coefficients of $\eta$. Thus equation
(\ref{round}) can be rewritten as
\bea
\frac1{2 \pi i} \oint \limits_{\infty} \frac{d \zeta}{\zeta^{m}}
\; \tilde\chi \ =\ \frac1{2 \pi i} \oint \limits_0 \frac{d
\zeta}{\zeta^{m+2-2n}} \;
\chi &-& \frac{1}{2 \pi i} \oint\limits_C \frac{d \zeta}{\zeta^m}
f' (\eta,\zeta) \nonumber \\
&-& \frac1{2 \pi i} \oint\limits_\infty \frac{d \zeta}{\zeta^m}
f(\eta,\zeta)\ .
\label{compare}
\eea
To make the notation more compact we combine the last two terms into
one integral of a new function $\hat f (\eta,\zeta)$.
For every particular example it will be clear which is the
contribution coming from the singularities of $\chi$, and what kind of
contour is to be taken around these singularities.
\bea
\frac{1}{2 \pi i} \oint \limits_{\infty} \frac{d \zeta}{\zeta^m} \; \tilde
\chi = \frac{1}{2 \pi i} \oint \limits_{0} \frac{d \zeta}{\zeta^{m+2-2n}} \;
\chi - \frac{1}{2 \pi i} \oint \limits_{C} \frac{d \zeta}{\zeta^m} \; \hat
f (\eta,\zeta)\ .
\label{hmm}
\eea

It is convenient to intoduce a new function $G (\eta,\zeta)$ such
that $\partial G /\partial \eta = \zeta^{2-2n}\hat f$, as well as the
following integral of $G$
\bea
 F = \frac{1}{2 \pi i} \oint \limits_{C} \frac{d \zeta}{\zeta^2} \;
G (\eta,\zeta)\ .
\eea
Here $F$ is a function of the coefficients of $\eta (\zeta)$, and is
precisely the function of section two. We also have
\bea
F_{w_j} = \frac{1}{2 \pi i} \oint \limits_{C} \frac{d \zeta}
{\zeta^{2n-j}} \hat f (\eta,\zeta)\ .
\eea
A comparison with (\ref{hmm}) and the expansion series shows that
$F_{w_{a'}} = 0 \; \text{for} \;2 \leq a' \leq 2n-2$, $F_z = t, \; F_v =
u,\; F_{\bar v} = (-1)^{n} \tilde u, \; F_{\bar z} = (-1)^{n+1} \tilde
t$. For the K\"ahler potential to be real the function $F$ should be
real and it follows that $\tilde t = (-1)^{n+1} \bar t, \; \tilde u =
(-1)^n \bar u, \ etc.$

{}From (\ref{holform}), the K\"ahler form is the coefficient of $\zeta$ in
the expansion of the $\I(\zeta)$ holomorphic form $\omega_h$,
and (\ref{omega}) then implies
\bea
\omega_1 = d z \wedge d t + d v \wedge d u = d z \wedge d (F_z) +
d v \wedge d u\ .
\label{almost}
\eea
Defining a new function
\bea
K (z,\bar z,u,\bar u) = F (z,\bar z,v,\bar v,w_{a'}) -
u v -\bar u \bar v\ ,
\label{pot}
\eea
and taking into account the relations already inferred,
\bea
F_v = u\ , \ \ F_{\bar v} = \bar u\ , \ \text{and}\ F_{w_{a'}} = 0\ ;
\ \ 2 \leq a \leq 2n-2\ ,
\eea
we derive the following relations
\begin{align}
K_z &= F_z + F_v \frac{\partial v}{\partial z} + F_{\bar v}
\frac{\partial
\bar v}{\partial z} + F_{w_{a'}}
\frac{\partial w_{a'}}{\partial z} - u \frac{
\partial v}{\partial z} - \bar u
\frac{\partial \bar v}{\partial z} = F_z \ ,\\
d v &= - d K_u\ .
\end{align}
Substitution into (\ref{almost}) gives
\begin{align}
\begin{split}
\omega_1 &= d z \wedge d (K_z) - d (K_u) \wedge d u = \\
&= K_{z \bar z} d z \wedge d \bar z + K_{z \bar u} d z \wedge d \bar u +
K_{u \bar z} d u \wedge d \bar z + K_{u \bar u} d u \wedge d \bar u\ ,
\end{split}
\end{align}
and so $\omega_1$ is the K\"ahler form for a metric with
K\"ahler potential $K$ (\ref{pot}). This ends our formal construction,
and explains the generalized Legendre transform construction in
a twistor space setting. We now consider examples.

\section{$A_k \; ALE$ metrics reconsidered}

As a first illustration of the above construction, we reinterpret the
twistor space construction of the $A_k$ asymptotically locally euclidean
metrics \cite{HAK} by the Legendre transform. The twistor space $\T$ of
the $A_k \; ALE$ metrics can be described as the zero locus\footnote{This
is true up to resolution of some singularities, but this subtlety will not
influence our discussion.} of the following polynomial:
\bea
x (\zeta) y (\zeta) - \prod_{i=1}^{k+1} (\eta (\zeta) - a_i (\zeta)) =
0\ ,
\label{ak}
\eea
where,
\bea
x (\zeta) \in \G(\oo (k+1))\ , \ \ y (\zeta) \in\G( \oo (k+1))\ ,\ 
\ \eta (\zeta) \in \G(\oo (2))\ ,
\eea
$a_i$ are fixed real (with respect to the real structure) sections of $\oo
(2)$ that parametrize the moduli space of metrics, $\eta$ is also real
and its coefficients are determined by 	the coordinates on the manifold.
The real structure exchanges $x$ and $y$. We write $\eta$ and
$a_i$ explicitly as:
\bea
\eta = z + v \zeta - \bar z \zeta^2, \quad a_i = p_i + q_i \zeta -
\bar p_i \zeta^2\ .
\eea
Guided by the real structure, Hitchin \cite{HAK} solves (\ref{ak}) 
as follows. Let
\bea
\eta - a_i = (- \bar z + \bar p_i) (\zeta - \alpha_i)
(\zeta + 1/{\bar \alpha_i})\ ,\\
\nonumber\\
\alpha_i = \frac{(v - q_i) - \sqrt{(v - q_i)^2 + 4 |z - p_i|^2}}
{2 (\bar z - \bar p_i)}\ .
\eea
Then set
\bea
x = A \prod_{i=1}^{k+1} (\zeta - \alpha_i)\ , \ \
y = B \prod_{i=1}^{k+1} (\zeta + 1/\bar \alpha_i)\ ,
\eea
where $A B = \prod (- \bar z + \bar p_i)$ and $ A \bar A = \prod ((v - q_i) +
\sqrt{(v - q_i)^2 + 4 |z - p_i|^2})$.

The holomorphic form is given by
\bea
\omega_h = d \eta \wedge \left( \frac{d x}{x} - \frac{d y}{y} \right)\ .
\eea
Thus the second $\I ({\zeta})$-holomorphic coordinate is
\bea
\chi = \ln (\frac{x}{y})\ .
\eea
 These considerations are all valid on the patch $\zeta \in U$. The
corresponding quantities on the other patch $\tilde U$ are
\bea
\tilde \eta = \zeta^{-2} \eta\ , \ \ \tilde x = \zeta^{-(k+1)} x\ ,\ \
\tilde y = \zeta^{-(k+1)} y\ .
\eea
These relations imply
\bea
\tilde \chi = \chi
\eea
and the function $f(\eta,\zeta)$ in (\ref{important}) vanishes. By
comparing the expansions of $\chi$ and $\tilde \chi$ as in (\ref{hmm}) we
read off the function $\hat f(\eta, \zeta)$ and the contour of integration
\bea
 \oint \limits_{C} \frac{d \zeta}{\zeta^m} \; \hat f(\eta,\zeta) =
\left( \oint \limits_{0} - \oint \limits_{\infty} \right) \frac{d \zeta}
{\zeta^m} \; \ln (\frac{x}{y}).
\label{kala}
\eea
The pairs of zeroes of $\eta - a_i$ are zeroes of $x$ and $y$ in the
logarithm in the right hand side above, and the two contours around $0$
and $\infty$ can be deformed to $k+1$ contours around the pairs of
roots of $\eta - a_i$.
Equation (\ref{kala}) can be rewritten as
\bea
\oint \limits_{C} \frac{d \zeta}{\zeta^m} \; \hat f (\eta,\zeta) =
\oint \limits_{C} \frac{d \zeta}{\zeta^m} \; \sum \limits_{i=1}^{k+1}
\ln (\eta - a_i)\ ,
\eea
where the contour $C$ circles each root with the opposite
orientation (after using the identity in Figure 1).
\begin{figure}[ht]
\unitlength 2.5em
\begin{picture}(15,2.25)(0,-1)
\centerline{
\epsfxsize=19em
\epsfclipon
\epsfbox{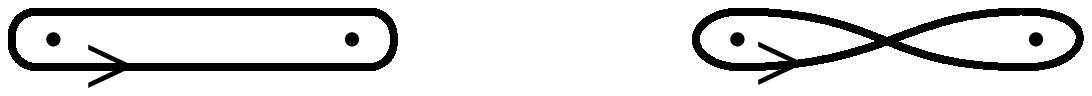}}
\put(-10,1.1){$C_1$:}
\put(-5.5,1.1){$C_2$:}
\put(-11,-.5){$\displaystyle\oint \limits_{C_1}d \zeta\, f(\zeta)\,
\ln \left( \frac{\zeta - a}{\zeta - b} 
\right)$}
\put(-7,-.5){$=$}
\put(-6.5,-.5){$\displaystyle\oint \limits_{C_2} d \zeta\, f(\zeta)\,
\ln \left((\zeta - a)(\zeta -b)\right)$}
\end{picture}
\caption{An identity between integrals.}
\label{pesho}
\end{figure}
Integrating $\hat f (\eta,\zeta)$ once with
respect to $\eta$ we obtain the function $G (\eta,\zeta)$
\bea
G (\eta,\zeta) = \sum \limits_{i=1}^{k+1} (\eta - a_i)
(\ln (\eta - a_i) -1)\ ,
\eea
in agreement with \cite{hklr}. For future reference, we also give the
function $F$ for the Taub-NUT metric:
\bea\label{TN}
F=\frac1{2\pi i}\oint d\zeta\, \left( \frac{\eta^2}{\zeta^3} +
m\frac{\eta(\ln\eta-1)}{\zeta^2}\right)\ ,
\eea
where $m$ is called the mass parameter.

\section{The Atiyah-Hitchin metric}

As another illustration of the generalized Legenedre transform
construction, we consider the Atiyah-Hitchin metric on the moduli space of
two centered $SU(2)$ monopoles. The metric is hyperk\"ahler and known
explicitely \cite{AH}. The twistor description is also known \cite{AH} and
the twistor space has a holomorphic projection on the bundle $\oo (4)$.
Thus one of the complex coordinates of the metric has an expansion
\bea
\eta (\zeta) = z + v \zeta + w \zeta^2 - \bar v \zeta^3 +
\bar z \zeta^4\ .
\label{quartic}
\eea
Due to the reality condition, the roots of the quartic (\ref{quartic})
come in two pairs of the type $\lambda, -1/\bar \lambda$. It is
convenient to write the quartic in terms of its roots and a scale
factor $c$:
\bea
\eta (\zeta) = c (\zeta - \alpha) (\zeta - \beta) (\bar \alpha \zeta + 1)
(\bar \beta \zeta + 1)\ .
\eea
A starting point for the generalized Legendre transform construction
is the function $G(\eta, \zeta)$. Asymptotically, the Atiyah-Hitchin 
metric approaches the Taub-NUT metric, and 
$\eta_{AH}\to\eta^2_{Taub-NUT}$. Comparing to (\ref{TN}), we make
the ansatz that the function $F$ for the Atiyah-Hitchin metric is
given by
\bea
F = - \frac{1}{2 \pi i} \oint \limits_{0}
\frac{d \zeta}{\zeta^3}
\; \eta + \oint \limits_{C} \frac{d \zeta}{\zeta^2} \; \sqrt{\eta}\ .
\label{ahf}
\eea
The contour $C$ encloses all four roots of $\eta$, and the branch cuts are
chosen to run from $\alpha$ to $-1/\bar\beta$ and from $\beta$ to
$-1/\bar\alpha$. The first step of the proof that (\ref{ahf}) does
indeed give the Atiyah-Hitchin metric is to solve the equation
\bea
F_w = - 1 + \oint \limits_{C} \frac{d \zeta}{\sqrt{\eta}} = 0\ ,
\label{fw}
\eea
for $w$, or equivalently, to find the four parameter family of twistor
lines inside the five real parameter family of quartics.
The integral in (\ref{fw}) is a complete elliptic integral of the
first kind; we transform it to Legendre normal form (\cite{BAT}, p. 307).
\bea
F_w = -1 + \frac{4}{\sqrt{c (1 + \alpha \bar \alpha)
(1 + \beta \bar \beta)}} \;
\int \limits_{0}^{1} \frac{d \zeta}{\sqrt{(1 - \zeta^2)(1 - k^2
\zeta^2)}} = 0\ ,
\label{ffw}
\eea
where the modulus is
\bea
k^2 = \frac{(1+ \alpha \bar \beta)(1 + \beta \bar \alpha)}{(1 + \alpha
\bar \alpha)(1 + \beta \bar \beta)}\ .
\eea
It does not seem possible to solve for $w$ explicitly from (\ref{ffw}),
but instead we can solve for the scale in terms of the roots
\bea
c = \frac{16 K^2 (k)}{(1 + \alpha \bar \alpha)(1 + \beta \bar \beta)}\ ,
\label{scale}
\eea
where $K (k)$ is the Legendre complete integral of first kind. To make contact
with the known form of the Atiyah-Hitchin metric we also reparametrize the
roots of the quartic (or equivalently the coefficients) using new variables
$\theta, \phi, \psi$, and then the restriction (\ref{fw}) on the
coefficients of the quartic (\ref{quartic}) is solved as follows
\begin{align}
\begin{split}
z &= 2 \exp^{2 i \phi} ( \cos(2 \psi) (1 + \cos^2(\theta)) +\\
&\qquad \qquad \qquad \quad
+ 2 i \sin(2 \psi) \cos(\theta) + (2 k^2 - 1) \sin^2(\theta)) K^2(k)\ ,
\end{split}\\
\begin{split}
v &= 8 \exp^{i \phi} \sin(\theta) ( \sin(2 \psi) -\\
&\qquad \qquad \qquad \quad
- i \cos(2 \psi) \cos(\theta) + i (2 k^2 - 1) \cos(\theta)) K^2(k)\ ,
\end{split}\\
w &= 4 (- 3 \cos(2 \psi) \sin^2(\theta) + (2 k^2 - 1) (1 - 3 \cos^2(\theta)))
K^2(k)\ .
\label{coeff}
\end{align}
To check that (\ref{coeff}) solves (\ref{scale}), we found the
roots of the quartic with these coefficients; this also determines the
scale, and then the direct substitution shows that (\ref{scale}) is
satisfied. The expressions for the roots are quite long and not very
illuminating, so we do not include them here. To clarify the particular
form of the coefficients (\ref{coeff}), notice that they and their complex
conjugates $\bar v,\; \bar z$ form a five dimensional
representation of the $SU(2)$ that rotates the sphere of complex
structures (acts on $\zeta$ by fractional transformations).
The vector fields generating this action have the standard
form \cite{GNS}:
\begin{align}
{\cal L}_3 &= -i \frac{\partial}{\partial \phi} \nonumber \\
{\cal L}_{+} &= e^{i \phi} \left( i \frac{\partial}{\partial \theta} +
\frac{1}{\sin(\theta)} \frac{\partial}{\partial \psi} - \cot(\theta)
\frac{\partial}{\partial \phi} \right) \label{vector} \\
 {\cal L}_{-} &= e^{-i \phi} \left( - i \frac{\partial}{\partial \theta} +
\frac{1}{\sin(\theta)} \frac{\partial}{\partial \psi} - \cot(\theta)
\frac{\partial}{\partial \phi} \right)\ . \nonumber
\end{align}
Now the four variables $k, \theta, \phi, \psi$ select a four parameter
family of quartics and parametrize the Atiyah-Hitchin manifold $\M$.

Next we can substitute the function $F$ into the formula for the
K\"ahler potential (\ref{pot}), compute the metric and compare with the
known form of the Atiyah-Hitchin metric:
\bea
 d s^2 = d \kappa^2 + a^2 \sigma_1^2 + b^2 \sigma_2^2 + c^2
\sigma_3^2\ ,
\eea
where
\bea
a b = - 2 E K + 2 {k'}^2 K^2\ ,\ \ b c = - 2 E K\ , \ \ c a = - 2 E K + 2
K^2\ ,
\eea
and $ k^2 + {k'}^2 = 1, \; K'(k) = K(k'), \; d \kappa = -(abc) d (K'/\pi
K) $, and $E$ is the complete elliptic integral of second kind.
Here, $\sigma_1,\; \sigma_2, \; \sigma_3$  are the forms invariant
with respect to the $SU(2)$ generated by the  vector fields
(\ref{vector}). This direct computation, however, is quite messy and
we instead checked that the complex structure $\I$ and the K\"ahler form
$\omega_1$ as computed from the generalized Legendre transform
construction coincide with the known ones: $z$ is clearly a
holomorphic coordinate and a lengthy calculation shows that
$u = F_v$ is holomorphic with respect to the known complex
structure \I\ as well. The K\"ahler form
$\omega_1 = dz \wedge d (F_z) + d v\wedge d (F_v)$ is easy to compute.
In a basis $d \kappa, d \theta, d\phi, d\psi$, its components are:
\bea
\omega_1 [1,2] &=& (b-a) \sin(\psi) \cos(\psi) \sin(\theta),
\nonumber \\
\omega_1 [1,3] &=& a \sin^2(\theta) \cos^2(\psi) + b \sin^2(\theta)
\sin^2(\psi) + c \cos^2(\theta) \nonumber \\
\omega_1 [1,4] &=& c \cos(\theta) \nonumber \\
\omega_1 [2,3] &=& (- ab + ac \sin^2(\psi) + bc \cos^2(\psi))
\sin(\theta)
\cos(\theta) \nonumber \\
\omega_1 [2,4] &=& c (a \sin^2(\psi) + b \cos^2(\psi)) \sin(\theta)
\nonumber \\
\omega_1 [3,4] &=& (b-a) c \sin(\psi) \cos(\psi) \sin^2(\theta).
\nonumber
\eea
This is precisely the K\"ahler form of the Atiyah-Hitchin metric.

\section{More examples, conjectures and conclusions.}

Given a function $G (\eta,\zeta)$, the (generalized) Legendre  transform
construction produces a hyperk\"ahler metric. Unfortunately, this
technique is purely local; it is not clear how to address the issue  of
geodesic completeness in general. We notice, however, that for many known
complete metrics the function $G (\eta,\zeta)$ has a very simple form: it
is a superposition of the same terms at different mass points. For
example, in the class of metrics with triholomorphic isometry (so that the
twistor space projects onto $\oo (2)$) the $A_k \; ALE$-metrics are
constructed out of a function $G$ that is a superposition at different
``mass'' points of the following basic block:
$G_0 = \eta (\ln (\eta) -1)$. A function $G$ that consists of just one
term like this gives flat space. Another form of $G$ that corresponds to
flat space is $G_{nut} = \eta^2/\zeta$. It is known
\cite{hklr} that adding a $G_{NUT}$-term to superposition of $G_0$ terms
gives a complete metric of Taub-NUT type. In particular, for $G = G_0 +
mG_{NUT}$, the Legendre transform construction gives the Taub-NUT metric
with mass parameter $m$ (\ref{TN}).

Next let us consider metrics whose twistor space projects on $\oo (4)$.
The function $G$ in the case considered section six consists of two parts
$\tilde G_{NUT} = \eta /\zeta$ and $\tilde G_0 = \sqrt{\eta}$. The
Atiyah-Hitchin metric asymptotically approaches the
Taub-NUT metric (for $m=-1$). In this limit $\eta_{AH} = \eta^2
_{Taub-NUT}$, and the twoparts of the function $G$ go to $G_0$ and
$G_{NUT}$ of the Taub-NUT metric (the $\ln(\eta)$ in $G_0$ serves merely to 
define the contour of integration, and doesn't appear in the Atiyah-Hitchin 
case). We conjecture that superimposing terms
of the type $\tilde G_0$ and $\tilde G_{NUT}$ gives new complete metrics.

There are also descriptions of flat space whose twistor space naturally
projects on $\oo (4)$. These have not been used to construct complete
hyperk\"ahler metrics, but are obvious candidate ingredients.  They are
$G=\eta^2/\zeta^3$ and $G=\eta (\ln (\eta) -1)/\zeta$ where in the latter
case, the contour for the $\ln(\eta)$ term encloses pairs of the roots of
$\eta$ as in contour $C_2$ of Figure 1.

As another example, we conjecture that the expression
\bea
G(\eta,\zeta)
=\sum_{i=1}^{k} ( (\sqrt{\eta} - a_i) \ln(\sqrt{\eta} - a_i) +
(\sqrt{\eta} + a_i) \ln (\sqrt{\eta} + a_i))
\eea
is related to $D_k \; ALE$-metrics \cite{iti}. Work on this conjecture is 
in progress \cite{iti}. The addition of
the term $\tilde G_{NUT}$ is interesting, and in
particular the metric with $k=2$ has been considered by Hitchin as a
degenerate case of $K3$ \cite{hit}.

Finally, we make a few comments about higher dimensional
hyperk\"ahler metrics. In the $\oo (2)$ case, examples based on
superpositions of $G_0$ and $G_{NUT}$ involving several independent
multiplets $\eta_i$ were given in \cite{lr1}. Recently, Gibbons and
Manton have used such a metric to describe the asymptotic form of the
metric on multi-monopole moduli spaces \cite{gm}; it is tempting to
speculate that a similar expression using our description of the
Atiyah-Hitchin metric may give at least a better approximation.

\bigbreak\medskip

{\Large\bf Acknowledgements}

\vskip .2in

We are indebted to Claude LeBrun for the numerous discussions from which
we benefited greatly. We also thank D.\ Jatkar for
discussions in the early stages of this project, and J.\ de Boer
for carefully reading the manuscript. 
This work is supported in part by NSF Grant no. PHY9309888.
\vspace{\medskipamount}

\end{document}